\documentclass[twocolumn]{revtex4-2}

\usepackage{natbib,bm}
\usepackage{amssymb,amsmath}

\usepackage[mathscr]{eucal} 
\usepackage{url}
\usepackage{colortbl,array,xcolor,color}
\usepackage[dvipdfmx]{graphicx}

\begin{document}

\title{Quasilinear flux model consistent with gyrokinetic ordering}

\author{O. Yamagishi\thanks{Corresponding author: yamagishi.osamu@nifs.ac.jp}}
\affiliation{National Institute for Fusion Science, Toki, Gifu, 509-5292, Japan}

\author{G. Watanabe}
\affiliation{School of Physics and Zhejiang Institute of Modern Physics, Zhejiang University, Hangzhou, Zhejiang 310027, China}

\begin{abstract}
  We propose a quasilinear (QL) flux model in which the saturation amplitude is uniquely determined
  using multiscale gyrokinetic ordering relations.
  The model is fully self-contained within a linear framework and does not rely on calibration
  against nonlinear simulations or mixing-length estimates.
  The wavenumber-dependent flux is given in ion gyro-Bohm units with a weighting factor of $|k_\theta \rho_i|$,
  such that its area integral in the log-linear scale yields the total flux, as employed in multiscale simulations.
  In systems with comparable ion and electron temperature gradients,
  the QL ion energy flux reproduces nonlinear simulation results in both its wavenumber dependence and absolute magnitude.
  In contrast, the QL electron flux is predominantly generated at electron scales,
  indicating that the shift of electron-scale transport toward ion scales observed in nonlinear Gsimulations
  is not captured within the present linear framework.
  We argue that the relation $Q_i\sim Q_e$,
  obtained as a closed conclusion of the QL model,
  may be predictive of simulation results if the area-integrated flux is conserved in nonlinear energy cascade process.
\end{abstract}
\maketitle

\section{Introduction}

In magnetic confinement devices, microscopic turbulence drives radial transport of particles and energy~\cite{Horton99,Connor00},
thereby degrading confinement and making its accurate evaluation essential. Because turbulent fluxes are inherently nonlinear,
arising at second order in the fluctuations, quantitative predictions generally require nonlinear simulations.
However, such simulations remain computationally expensive.
Quasilinear (QL) flux models, which reconstruct second-order fluxes from linear calculations,
are therefore widely used to provide rapid insight into transport responses to parameter variations~\cite{Staebler24}.

QL flux models have evolved from early formulations~\cite{Rewoldt87, Romanelli88}
into increasingly sophisticated frameworks~\cite{Waltz97, Staebler05, Jenko05, Bourdelle07, Citrin12, Fransson22, Stephens24, Giacomin25}.
Beyond the intrinsic challenge of bridging fluid and kinetic descriptions~\cite{Waltz97, Staebler05},
much of the complexity arises from model-specific prescriptions for the saturation amplitude and, in certain cases,
from the treatment of the squared perpendicular wavenumber in mixing-length estimates introduced
from a dimensional-analysis perspective~\cite{Jenko05}.
In some implementations, this amplitude is calibrated using nonlinear gyrokinetic simulations,
leading to reported agreement with fully nonlinear results~\cite{Staebler24}.
However, such complexity can hinder independent verification,
making it difficult to assess whether the agreement reflects genuine predictive capability
or features inherent to the model construction.

Because QL flux models do not inherently guarantee quantitative agreement with nonlinear simulations,
it is desirable to adopt a formulation that remains closed within a linear framework without introducing unnecessary complexity.
Motivated by this perspective,
we propose a QL flux model based on a simple model for the saturation amplitude,
consistent with gyrokinetic ordering~\cite{BrizardHahm,Frieman82}.
The model is designed to incorporate multiscale gyrokinetic ordering,
motivated by multiscale simulation results~\cite{Gorler08,Gorler08no2}.
The resulting QL flux is expressed in ion gyro-Bohm units with an appropriate weighting factor,
such that the total flux is given by its area integral in a log-linear representation,
as commonly employed in multiscale simulations~\cite{Gorler08,Howard14}.
The results highlight both the
predictive capability and the intrinsic limitations of QL models based solely on gyrokinetic ordering,
and provide a transparent baseline for interpreting multiscale turbulent transport without
recourse to nonlinear calibration.

\section{Quasilinear flux model}\label{sec:QL}

In this work, we consider the linear gyrokinetic equation in the ballooning representation,
coupled with Poisson equation~\cite{Taylor68,Catto81,Rewoldt82,Frieman82},
in which the perturbed distribution function responds to perturbed electrostatic potential $\phi$.
The associated $\bm{E}\times\bm{B}$ drift is taken to be $\delta\bm{u}_E = - i \phi\bm{k}_\perp  \times \bm{B}/B^2$.
Here, $\bm{B}$ denotes the equilibrium magnetic field vector,
and the perpendicular wavenumber vector is defined as $\bm{k}_\perp = k_\alpha \nabla \alpha + k_r \nabla r$,
where $\alpha = \zeta - q \theta$ is the field-line label, with $q$ the safety factor in toroidal coordinates
$(r, \theta, \zeta)$~\cite{HazeltineMeiss}.
 The radial component of particle and energy fluxes in wavenumber space are then given by~\cite{Rewoldt87,Staebler24}
\begin{align}
\begin{split}
  &\Gamma_{a}^k=\biggr\langle \int d^3v \delta\bm{u}_E^\ast\cdot\nabla r \delta f_a \biggr\rangle
  =\frac{T}{eB_0}
 \biggr\langle-ik_\theta\frac{e\phi^\ast}{T}\delta n_a\biggr\rangle,\\
  &Q_{a}^k=\biggr\langle \int d^3v \delta\bm{u}_E^\ast\cdot\nabla r \frac{1}{2}m_av^2\delta f_a \biggr\rangle
  =\frac{T}{eB_0}\biggr\langle-ik_\theta\frac{e\phi^\ast}{T}\frac{3}{2}\delta p_a\biggr\rangle,
  \label{eq:QL_fluxes}
  \end{split}
\end{align}
where $\delta\bm{u}_E^\ast\cdot\nabla r=-(i\phi)^\ast \bm{k}_\perp\times\bm{B}/B^2\cdot\nabla r=-i\phi^\ast k_\theta/B_0$,
with the poloidal wavenumber $k_\theta=-k_\alpha B_0/(d\chi/dr)$
which represents a reciprocal of spatial scale of fluctuations in the poloidal direction, and $\chi$ is a flux function satisfying $\bm{B}=\nabla\alpha\times\nabla\chi$.
We take $k_\theta<0$ throughout.
The notation $\langle\cdot\rangle$ denotes the flux-surface average, $e=|e_e|$,
and $B_0=\langle B^2\rangle^{1/2}$. The superscript $^\ast$ denotes complex conjugation.
The perturbed density and pressure are defined as
$\delta n_a=\int \delta f_a d^3v$ and $\delta p_a=\int (m_av^2/3)\delta f_a d^3v$,
where $\delta f_a$ denotes the perturbed distribution function at the particle position.
In the QL model, these quantities are assumed to share the same wavenumber $(k_\theta,k_r)$ as the potential $\phi$.
A common temperature is defined as $T=\sum_a T_a n_a / \sum_a n_a$,
corresponding to the equilibrium temperature toward which the species temperatures $T_a$ relax
under Fokker-Planck collisions when initially unequal~\cite{Yamagishi24}.
Only the real parts of the fluxes are retained as physically relevant quantities.

It is well-known that gyrokinetic equation is derived under the ordering~\cite{BrizardHahm,Frieman82},
\begin{align}
\begin{split}
  &O(\epsilon)\sim \frac{\rho_b}{L}\sim \frac{\omega}{\Omega_b}
  \sim\frac{e\phi}{T},\\
  &O(1)\sim k_\perp\rho_b,
\label{eq:gyrokinetic_orderings}
\end{split}
\end{align}
where $\epsilon\ll 1$.
The subscript $b$ specifies whether we are considering the ion~($b=i$) or the electron~($b=e$) gyrokinetics,
and is distinguished from the species index $a$ used in the flux $Q_a^k$.
Here we retain only the ordering relations relevant to the present model.
The quantity $L$ denotes an equilibrium scale length (typically the perpendicular variation scale of the magnetic field,
$L \sim |d\ln B/dr|^{-1}\sim R_0$, the major radius),
and $\rho_b = v_{tb}/|\Omega_b|$ is the Larmor radius of species $b$,
with $\Omega_b = e_b B_0/m_b$ and $v_{tb} = \sqrt{T/m_b}$.
The frequency $\omega \sim \partial/\partial t$ characterizes the typical fluctuation time scale.

We now introduce the following model for the squared amplitude based on these orderings:
\begin{align}
\left|\frac{e\phi}{T}\frac{L}{\rho_b}\right|^2 \sim O(1) \sim \biggr|\frac{\omega}{\Omega_b}\frac{L}{\rho_b}\biggr|^n (|k_\theta \rho_b|)^m,
\label{eq:model1}
\end{align}
with $n=1$ and $m=-1$.
Setting $n=1$ leads to the choice that the squared saturation amplitude is proportional
to the mode frequency (in particular, the growth rate) to the first power,
which is widely used in the QL models including the mixing-length estimate~\cite{Staebler24}.
On the other hand, as a consequence of choosing $m=-1$,
it will be shown numerically later in Fig.\ref{fig:01} that the ordering combinations on the right-hand side indeed yield $O(1)$.
For ion-scale fluctuations, $|k_\theta \rho_i| \sim 1$,
the QL energy flux is then given by
\begin{align}
  Q_a^k&=\frac{T}{eB_0}\biggr(\frac{\rho_{i}}{L}\biggr)^2
  \frac{ -ik_\theta \langle (e\phi^\ast/T)(3 \delta p_a/2) \rangle}
       {\langle |e\phi/T|^2\rangle}\biggr\langle\biggr|\frac{e\phi}{T}\frac{L}{\rho_{i}}\biggr|^2\biggr\rangle\nonumber\\
       &=p\chi_{\textrm{gB}}^i\frac{1}{L}\biggr[
         \frac{-ik_\theta\rho_{i}\langle (e\phi^\ast/T)(3\delta p_a/2p)\rangle}{\langle|e\phi/T|^2\rangle}
         \biggr(\frac{\gamma}{\omega_{\ast i}}\biggr)\biggr].
       \label{eq:QL_flux1}
\end{align}
Here, we replace $|\omega| L/(\Omega_i \rho_i)$ with $\gamma/(v_{ti}/L)$,
assuming that the linear growth rate $\gamma$ is more relevant to the saturation amplitude
than the real frequency $\omega_r$ of the complex mode frequency $\omega = \omega_r + i\gamma$.
We also introduce the diamagnetic frequency $\omega_{\ast b} = |k_\theta \rho_b| v_{tb}/L = |k_\theta| T/(e B_0 L)$,
which is independent of particle species and can therefore be written as $\omega_{\ast b} = \omega_\ast$.
In addition, we define the representative pressure $p = n_e T$ and the gyro-Bohm diffusivity for species $b$,
\begin{align}
\chi_{\mathrm{gB}}^b = \frac{\rho_b}{L}\frac{T}{e B_0},
\label{eq:gyroBohm_diffusion}
\end{align}
such that the bracketed term in Eq.~(\ref{eq:QL_flux1}) corresponds to the QL flux normalized to gyro-Bohm units.
The corresponding particle flux $\Gamma_a^k$ is obtained by replacing
$3\delta p_a/(2p) \rightarrow \delta n_a/n_e$ and $p \chi_{\mathrm{gB}}^i \rightarrow n_e D_{\mathrm{gB}}^i$
with $D_{\mathrm{gB}}^i=\chi_{\mathrm{gB}}^i$.

The above formulation for ion-scale fluctuations,
typically associated with ion temperature gradient (ITG) modes,
can be extended to electron-scale fluctuations with $|k_\theta \rho_e| \sim O(1)$
for electron temperature gradient (ETG) modes.
Applying the same procedure using electron gyrokinetic ordering yields expressions
analogous to Eqs.~(\ref{eq:model1})-(\ref{eq:gyroBohm_diffusion}), with the ion mass replaced by the electron mass.
In fact, within linear theory, electron gyrokinetics with adiabatic ions and ion gyrokinetics
with adiabatic electrons are known to be isomorphic~\cite{Dorland00}.
That is, the frequencies and wavenumbers of ITG and ETG modes,
when normalized by $v_{tb}/L$ and $1/\rho_b$ (with $b = i, e$, respectively), are identical
up to the sign of the real frequency and the phase of the potential,
reflecting the charge difference.
Even in the more general case where both species exhibit non-adiabatic responses,
the normalized QL flux (i.e., the bracketed term in Eq.~(\ref{eq:QL_flux1})) remains of the same order for ITG and ETG.
This leads to the conclusion that the QL flux associated with ETG is smaller than that of ITG
by a factor of $O(\sqrt{m_e/m_i})$, reflecting the $\rho_b$ dependence in the gyro-Bohm diffusivity
in Eq.~(\ref{eq:gyroBohm_diffusion})~\cite{Dorland00}.

Early nonlinear simulations treated ion- and electron-scale gyrokinetic dynamics separately.
Ion-scale studies showed that zonal flows suppress transport and shift fluctuation spectra to lower wavenumbers,
reducing the nonlinear critical gradient relative to linear thresholds~\cite{Lin98,Dimits00,Rogers00}.
In contrast, electron-scale simulations predicted large ETG-driven transport, $\chi/\chi_{gB}^e \sim O(10$-$100)$,
due to weak zonal-flow regulation and the formation of streamers~\cite{Dorland00,Jenko00},
though this contradicted the expected ITG-ETG similarity.
Multiscale simulations resolving both ion and electron dynamics demonstrated
that the adiabatic-ion approximation overestimates ETG transport~\cite{Nevins07,Waltz07,Candy07}.
They further showed that cross-scale coupling suppresses high-$k_\perp$ ETG transport
while enhancing low-$k_\perp$ ITG/TEM transport, leading to a broadened spectrum spanning ion to electron scales.
Notably, both ion and electron fluxes can peak at ion scales~\cite{Gorler08,Howard14}.

These simulation results motivate consideration of the intermediate wavenumber range bridging ion and electron scales.
We assume a $|k_\perp|^{-2}$ scaling of the squared amplitude across scales,
expressed as $|e\phi/T|^2 |k_\theta L|^2 \sim O(1)$, which is partly motivated by simulation results~\cite{Gorler08no2}.
In multiscale systems,
It is necessary to specify whether the wavenumber is normalized to the Larmor radius of an arbitrary species $b$,
which may be an ion or an electron.
Accordingly, the amplitude can be written as $|e\phi/T|^2 (L/\rho_b)^2 |k_\theta \rho_b|^2 \sim O(1)$,
which we refer to as the $b$-species representation.
This form satisfies the gyrokinetic ordering in Eq.~(\ref{eq:gyrokinetic_orderings}) at both scales:
$|e\phi/T|^2 (L/\rho_i)^2 \sim O(1)$ at $|k_\theta \rho_i| = 1$ in the ion representation,
and $|e\phi/T|^2 (L/\rho_e)^2 \sim O(1)$ at $|k_\theta \rho_e| = 1$ in the electron representation.
The amplitude model in Eq.~(\ref{eq:model1}) is then modified for $b$-species representation as follows:
\begin{align}
  \biggr|\frac{e\phi}{T}\frac{L}{\rho_{b}}\biggr|^2\sim \frac{\gamma}{\omega_{\ast b}}\frac{1}{|k_\theta\rho_{b}|^2}.
  \label{eq:model2}
\end{align}
The corresponding QL flux for general wavenumber is obtained as,
\begin{align}
  Q_a^k&=
       p\chi_{\textrm{gB}}^b\frac{1}{L}\biggr[
         \frac{i\langle(e\phi^\ast/T)(3\delta p_a/2p)\rangle}{|k_\theta\rho_b|\langle|e\phi/T|^2\rangle}
         \biggr(\frac{\gamma}{\omega_{\ast b}}\biggr)\biggr]\nonumber\\
       &=p\chi_{\textrm{gB}}^i\frac{1}{L}\frac{1}{|k_\theta\rho_i|}\biggr[
         \frac{i\langle(e\phi^\ast/T)(3\delta p_a/2p)\rangle}{\langle|e\phi/T|^2\rangle}
         \biggr(\frac{\gamma}{\omega_{\ast b}}\biggr)\biggr],
       \label{eq:QL_flux2}
\end{align}
where we have used $k_\theta/|k_\theta| = -1$.
As noted above, $\omega_{\ast b}=\omega_\ast$ is independent of $b$,
so that the final expression, when normalized to the ion gyro-Bohm diffusion coefficient, is uniquely determined regardless
of the choice of $b$-species representation.

\section{Numerical results}\label{sec:numerical_results}

Here, we present results from linear analyses and the QL flux model based on the gyrokinetic-Poisson system~\cite{Taylor68,Catto81,Rewoldt82}.
We consider a collisionless plasma and retain only the perturbed scalar potential, $\phi$.
The calculations are performed using the \texttt{GOBLIN} code~\cite{Yamagishi04}, which employs an eigenvalue formulation~\cite{Rewoldt82}.

We first consider the well-established cyclone base case parameters~\cite{Dimits00}.
To this end, we incorporate the large-aspect-ratio circular tokamak model, known as the $\bar{s}$-$\bar{\alpha}$ model~\cite{Connor78},
into the gyrokinetic equations (see Ref.~\cite{Yamagishi18} for details). The parameters are the inverse aspect ratio $r/R_0=0.18$,
density and temperature gradients $R_0/L_{n_a}=2.2$ and $R_0/L_{T_a}=6.9$
with $L_X^{-1}=-d\ln X/dr$, safety factor $q=1.4$, magnetic shear $\bar{s}=0.78$,
and pressure gradient parameter $\bar{\alpha}=0$. We consider an electron-deuterium plasma with $m_i/m_e=3672$.
The wavenumber dependence of the linear growth rate $\gamma$ and real frequency $\omega_r$ is evaluated by setting $k_r=0$ and varying only $k_\theta$.
Three representative instabilities, ion temperature gradient mode (ITG), trapped electron mode (TEM), and electron temperature gradient mode (ETG),
can be identified. The ITG mode is characterized by a negative real frequency, while TEM and ETG exhibit similar real frequencies at ion scales.
They are distinguished by the fact that
the ETG mode has a growth rate peak at the electron scale,
and therefore the associated potential is spread along magnetic field lines at the ion scale with $|k_\theta\rho_i|\sim 1$,
whereas that for TEM exhibits a more localized structure.
We observe that the frequencies of all three instabilities, normalized to $\omega_\ast$ ($\equiv \omega_{\ast i} \equiv \omega_{\ast e}$),
remain $O(1)$ across the wavenumber range from ion to electron scales when plotted on a log-linear scale.
This contrasts with the use of a log-log plot for the frequencies of ITG-ETG modes normalized by $v_{ti}/R_0$~\cite{Howard14},
thereby supporting the validity of the amplitude model in Eq.(\ref{eq:model1}).
We also note that the combination $\gamma/\omega_\ast \propto \gamma/k_\theta$ was used in Refs.~\cite{Staebler17, Creely19} to construct their models.

\begin{figure}[htbp]
\begin{center}
\begin{minipage}{.45\linewidth}
  \includegraphics[width=\linewidth]{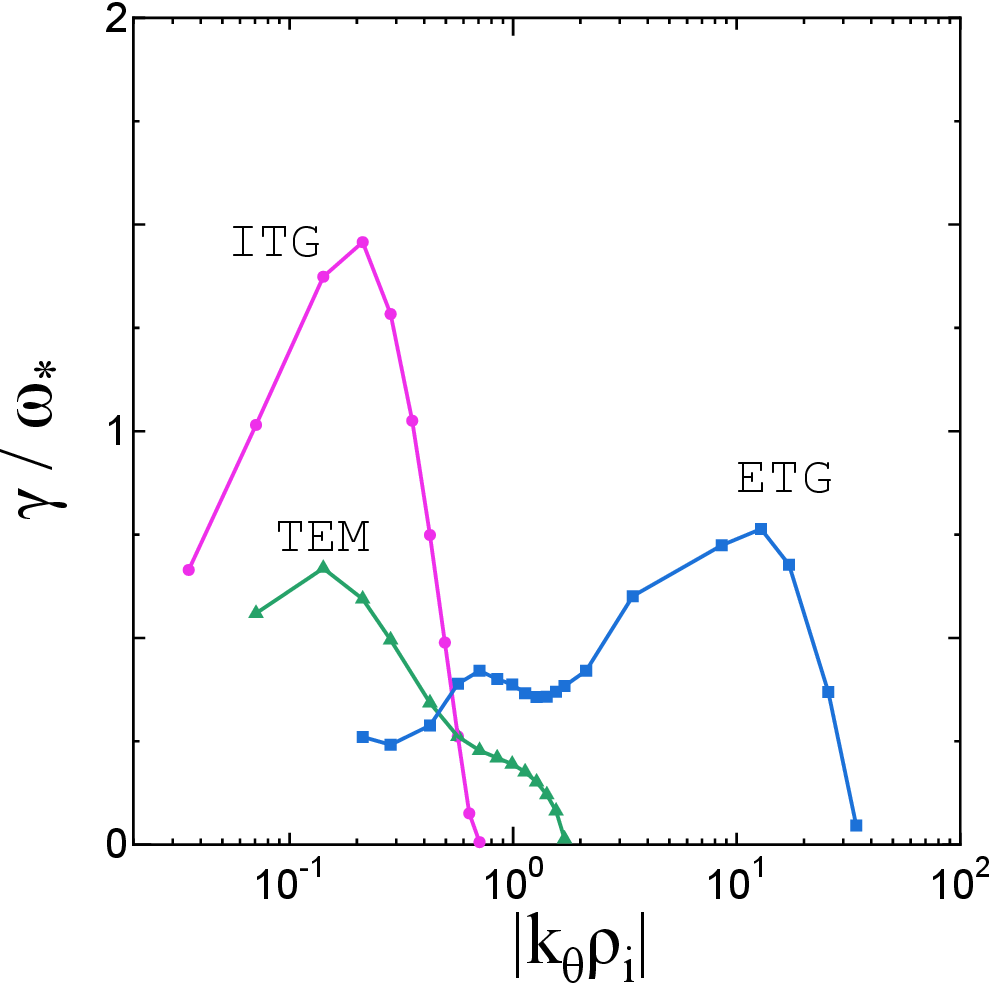}
\end{minipage}
\begin{minipage}{.45\linewidth}
  \includegraphics[width=\linewidth]{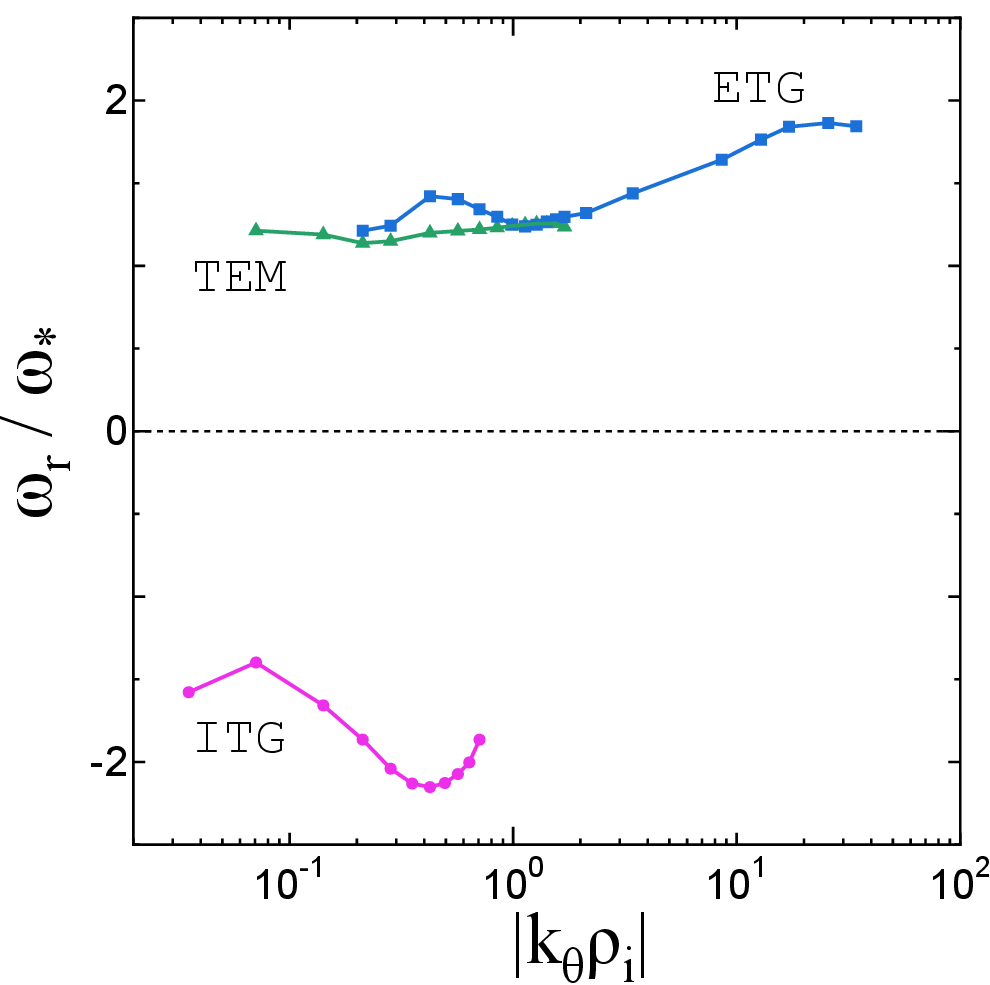}
\end{minipage}
\caption{Growth rates $\gamma$ (left) and real frequencies $\omega_r$ (right) as functions of the normalized wavenum
ber $|k_\theta \rho_i|$ for ITG (circles), TEM (triangles), and ETG (squares) modes in the Cyclone base case. The fr
equencies are normalized by the diamagnetic frequency, $\omega_\ast$, and shown on log-linear scales.}
\label{fig:01}
\end{center}
\end{figure}

\begin{figure}[htbp]
\begin{center}
\begin{minipage}{.45\linewidth}
  \includegraphics[width=\linewidth]{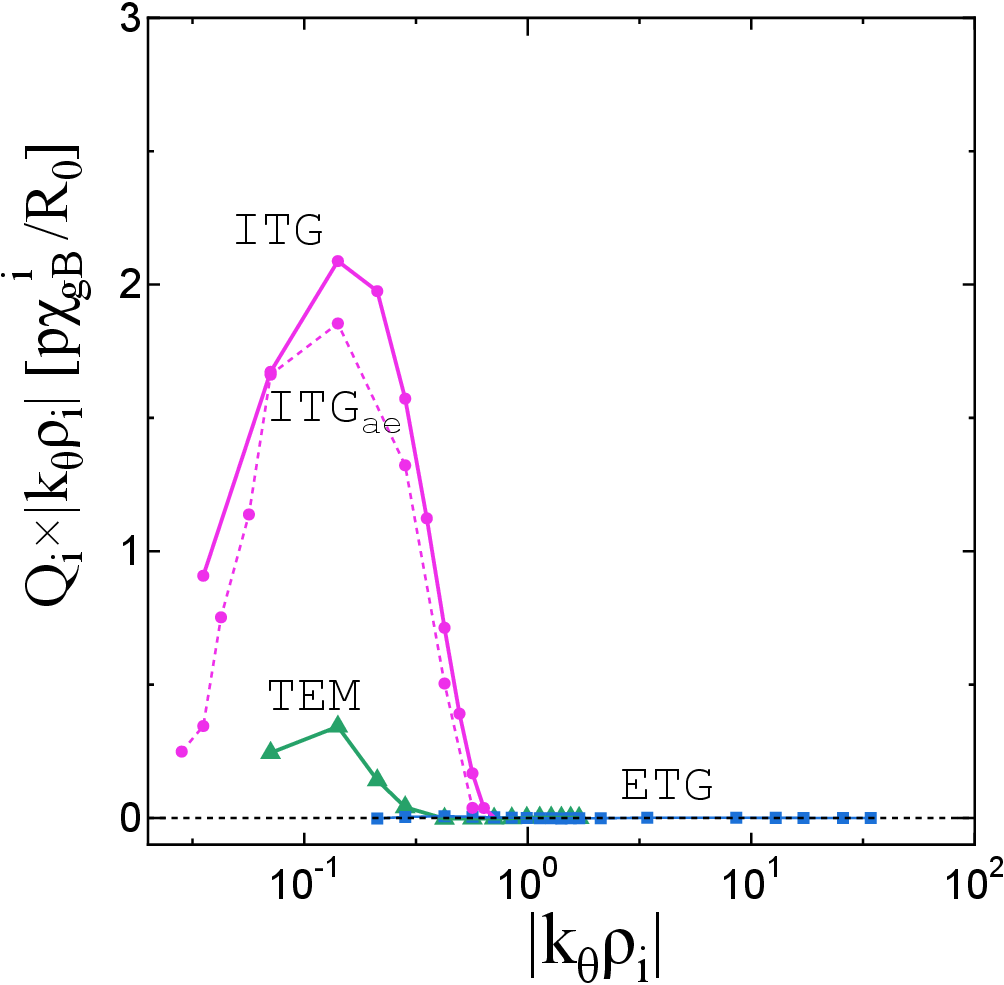}
\end{minipage}
\begin{minipage}{.45\linewidth}
  \includegraphics[width=\linewidth]{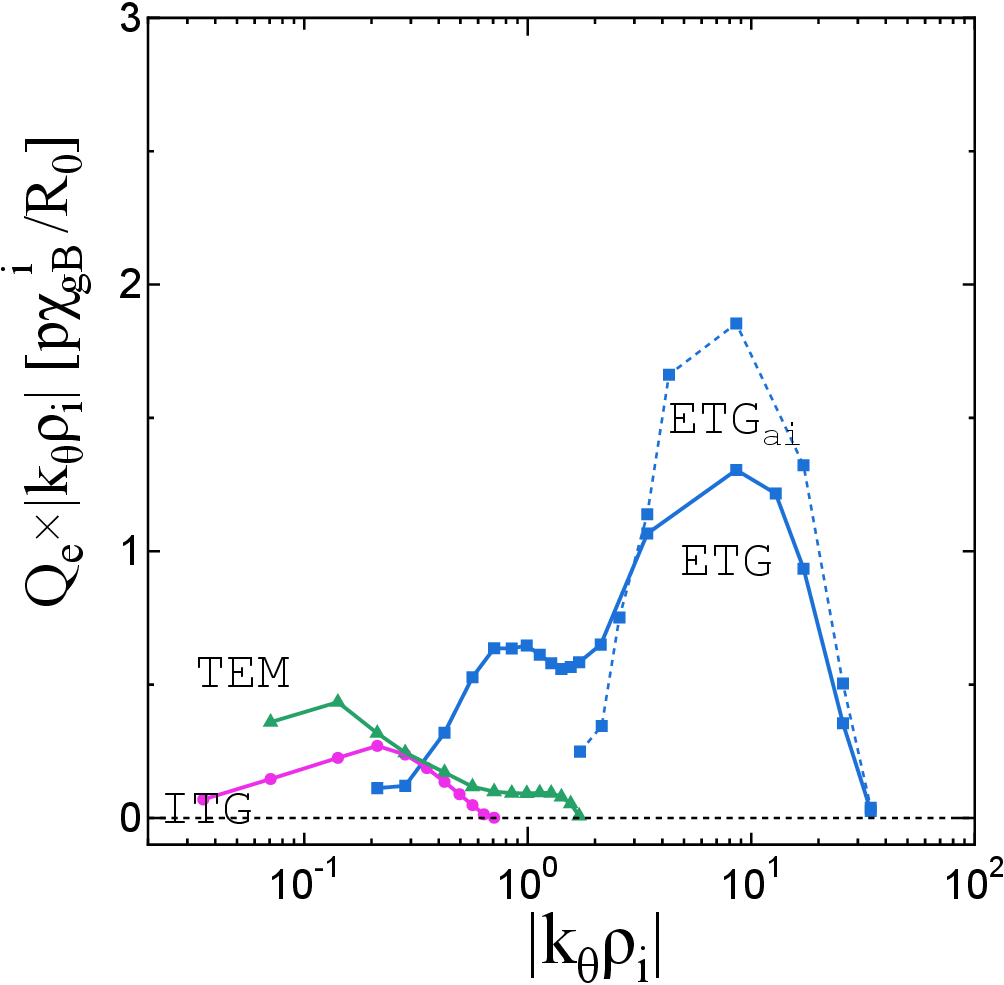}
\end{minipage}
\caption{QL ion (left) and electron (right) energy fluxes as functions of $|k_\theta \rho_i|$ for ITG (circles), TEM (triangles), and ETG (squares) modes in the Cyclone base case. The energy fluxes, weighted by $|k_\theta \rho_i|$ and expressed in ion gyro-Bohm units, are plotted on a log-linear scale.
For comparison, the results for ITG under the adiabatic electron~(ae) approximation and ETG under the adiabatic ion~(ai) approximation are also shown as dashed lines.}
\label{fig:02}
\end{center}
\end{figure}

The QL ion and electron energy fluxes corresponding to the instabilities in Fig.~\ref{fig:01} are shown in Fig.~\ref{fig:02}.
Following \cite{Gorler08}, we plot the QL flux weighted by $|k_\theta \rho_i|$ in ion gyro-Bohm units, with $L=R_0$ in Eq.~(\ref{eq:gyroBohm_diffusion}).
This choice is motivated by interpreting the total flux as the area integral of $Q_a^k$ over $|k_\theta \rho_i|$,
i.e., $Q_a \approx \int Q_a^k  d|k_\theta \rho_i| = \int Q_a^k |k_\theta \rho_i|  d(\log|k_\theta \rho_i|)$~\cite{Gorler08,Howard14}.
In the following text, for simplicity, we will denote the weighted flux $Q_a^k|k_\theta\rho_i|$ simply as $Q_a^k$.
The ion flux $Q_i^k$ is predominantly driven by ITG, and peaks at $|k_\theta \rho_i| \sim 0.2$,
consistent with typical multiscale simulations~\cite{Gorler08,Howard14}. Moreover, under the same normalization, its magnitude
is comparable to simulation results~\cite{Gorler08}.
On the other hand, the QL electron flux, mainly driven by ETG, peaks at the electron scale ($|k_\theta \rho_e| \sim 0.2$)
and is separated from the ion scale.
This contrasts with simulation results,
where the peak in the wavenumber spectrum of the electron flux appears at the ion scale ($|k_\theta \rho_i| < 1$)~\cite{Gorler08,Howard14}.

The appearance of a peak in the electron flux driven by ETG modes at ion scales, as observed in the simulations,
is clearly nonlinear in origin and is therefore attributed to a nonlinear energy cascade
and redistribution between species~\cite{Abel13}.
Although the QL model does not capture these nonlinear behaviors,
it is derived on an equal footing for ions and electrons based on gyrokinetic ordering.
It is therefore reasonable to expect that, if the model has predictive capability for the ion flux,
it should exhibit comparable predictive capability for the electron flux.
This expectation can be justified if the area-integrated flux predicted
by the QL model in the linear regime is approximately conserved during the nonlinear downshift in wavenumber space.
This assumption is plausible if the cascade is local in wavenumber space, so that,
in the intermediate range spanning ion and electron scales,
effects of energy dissipation at high wavenumber and nonlinearly driven flows at low wavenumber are negligible,
and if turbulence and equilibrium scales are well separated,
precluding radial energy exchange (e.g., turbulence spreading~\cite{Hahm04}) as in Ref.\cite{Abel13}.
The ansatz of the area-integral conservation and resulting predictability of the QL flux in the electron scale
are indirectly supported by a comparison between the ion heat flux driven by ITG under the adiabatic electron approximation,
$Q_i^{k(\mathrm{ITG,ae})}$, and the electron heat flux driven by ETG under the adiabatic ion approximation, $Q_e^{k(\mathrm{ETG,ai})}$,
shown by the dashed lines in Fig.~\ref{fig:02}.
Their area-integrated values are found to agree exactly,
indicating that the weighted fluxes for ITG(ae) and ETG(ai) are not isomorphic,
but rather translationally symmetric in $\log |k_\theta \rho_i|$ direction in this model.
The resulting electron thermal diffusivity satisfies $\chi_e = Q_e/(p/L) \sim \chi_\mathrm{gB}^i \gg \chi_\mathrm{gB}^e$,
consistent with early ETG simulations~\cite{Dorland00,Jenko00}.

\begin{figure}[htbp]
\begin{center}
\begin{minipage}{.43\linewidth}
  \includegraphics[width=\linewidth]{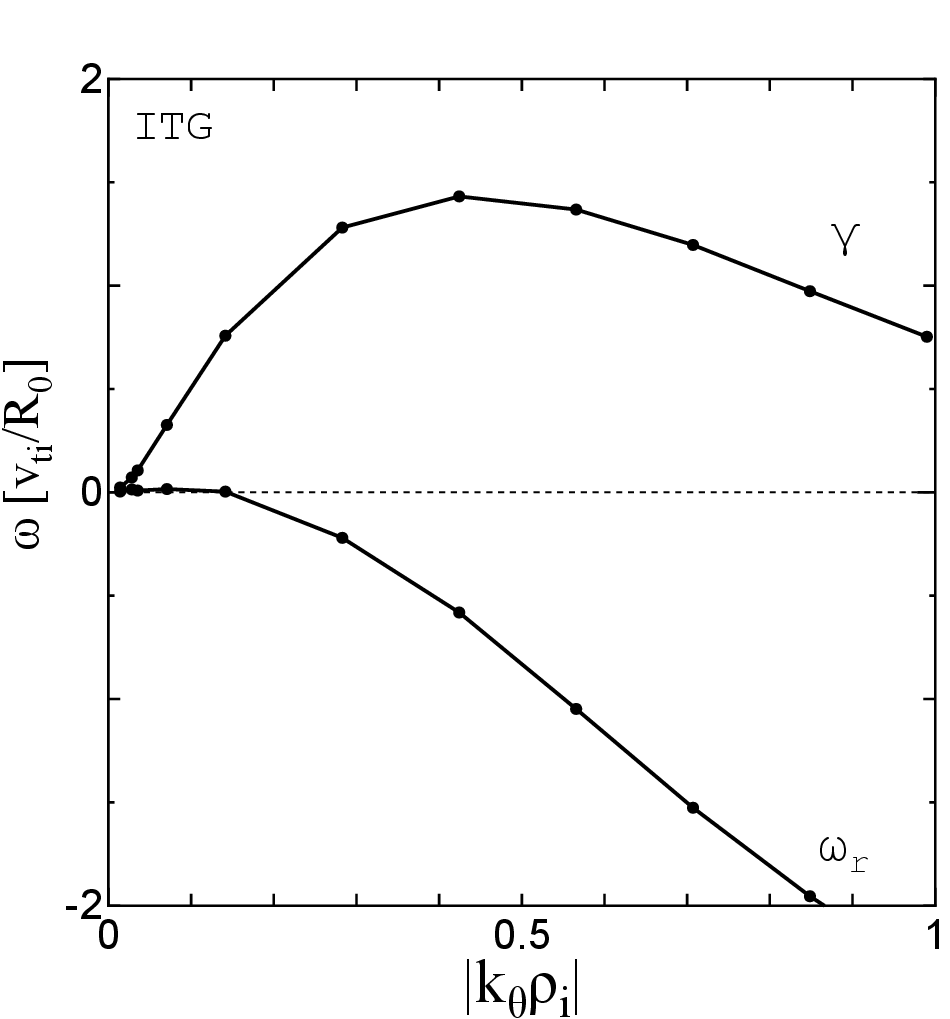}
\end{minipage}
\begin{minipage}{.43\linewidth}
  \includegraphics[width=\linewidth]{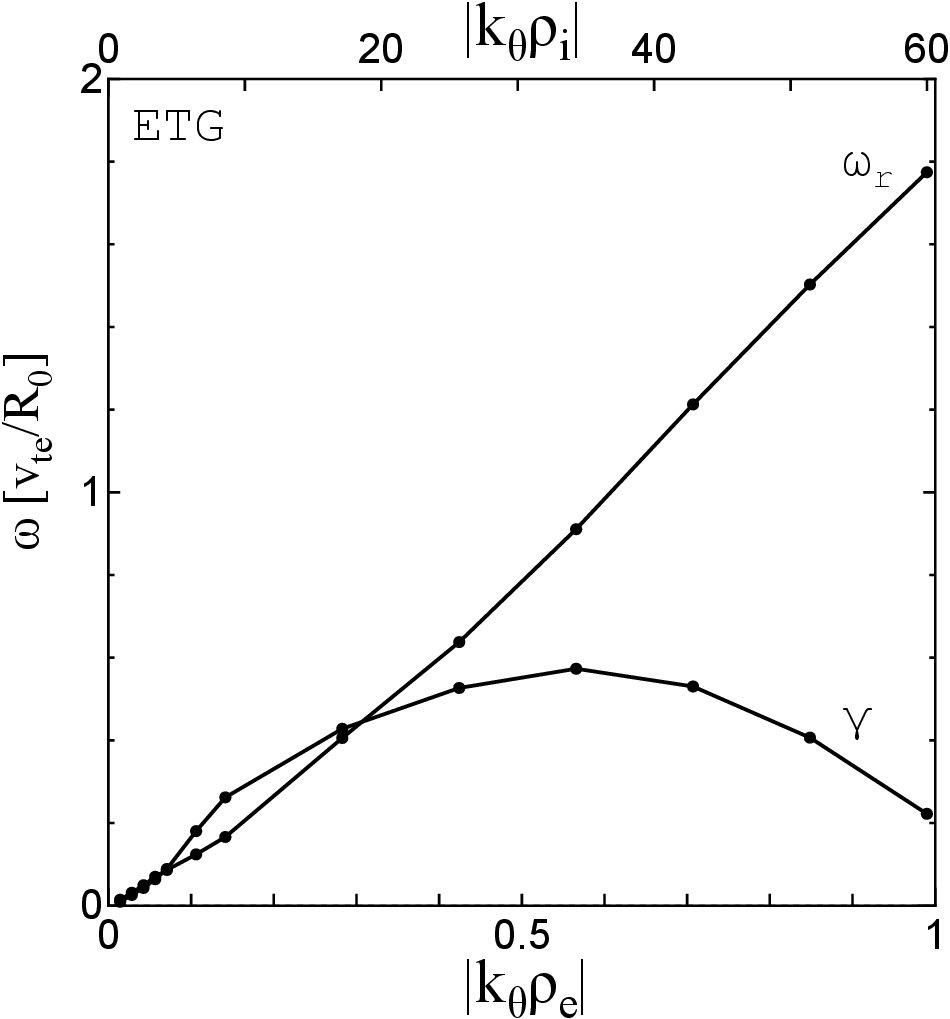}
\end{minipage}
\end{center}
\hspace{-20mm}
\\[80pt]
\begin{minipage}{.96\linewidth}
  \includegraphics[width=\linewidth]{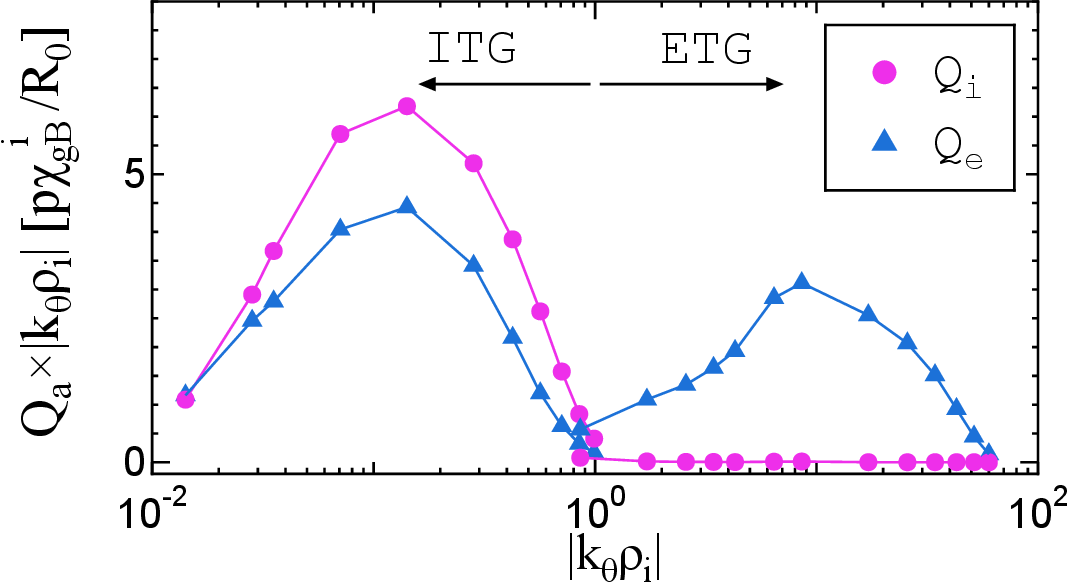}
\end{minipage}
\caption{(Top) Wavenumber dependence of frequencies (growth rate $\gamma$ and real frequency $\omega_r$) for ITG (left) and ETG (right) modes, where both frequencies and wavenumbers are normalized by the corresponding thermal frequency, $v_{ta}/R_0$, and inverse Larmor radius, $\rho_a^{-1}$.
  (Bottom) Corresponding QL energy fluxes as functions of $|k_\theta \rho_i|$, where the energy flux, weighted by $|k_\theta \rho_i|$ and expressed in ion gyro-Bohm units, is shown on a log-linear scale.}
\label{fig:03}
\end{figure}

Next, we consider a circular tokamak equilibrium with $R_0/r_\textrm{edge}=3$,
using an MHD equilibrium obtained numerically with \texttt{VMEC}~\cite{Hirshman83}.
In this test case, at the normalized minor radius $\rho=r/r_\textrm{edge}=0.7$,
the local parameters are $q\simeq1.73$, $\bar{s}\simeq1.7$, $\bar{\alpha}\simeq0.14$, $\beta_a\simeq0.0015$, $R_0/L_{T_a}\simeq10.4$,
and $R_0/L_{n_a}\simeq5$ for $a=i,e$.
The top panels of Fig.~\ref{fig:03} show the linear growth rates and real frequencies of the ITG and ETG modes,
normalized by $v_{ta}/R_0$, which peak at ion and electron scales, respectively.
The corresponding QL energy fluxes are shown in the bottom panel.
A significant fraction of $Q_e^k$ is found to be driven by the ITG mode,
indicating a substantial deviation from the $\bar{s}$-$\bar{\alpha}$ model.

\begin{figure}[htbp]
\begin{center}
  \begin{minipage}{.45\linewidth}
    \includegraphics[width=\linewidth]{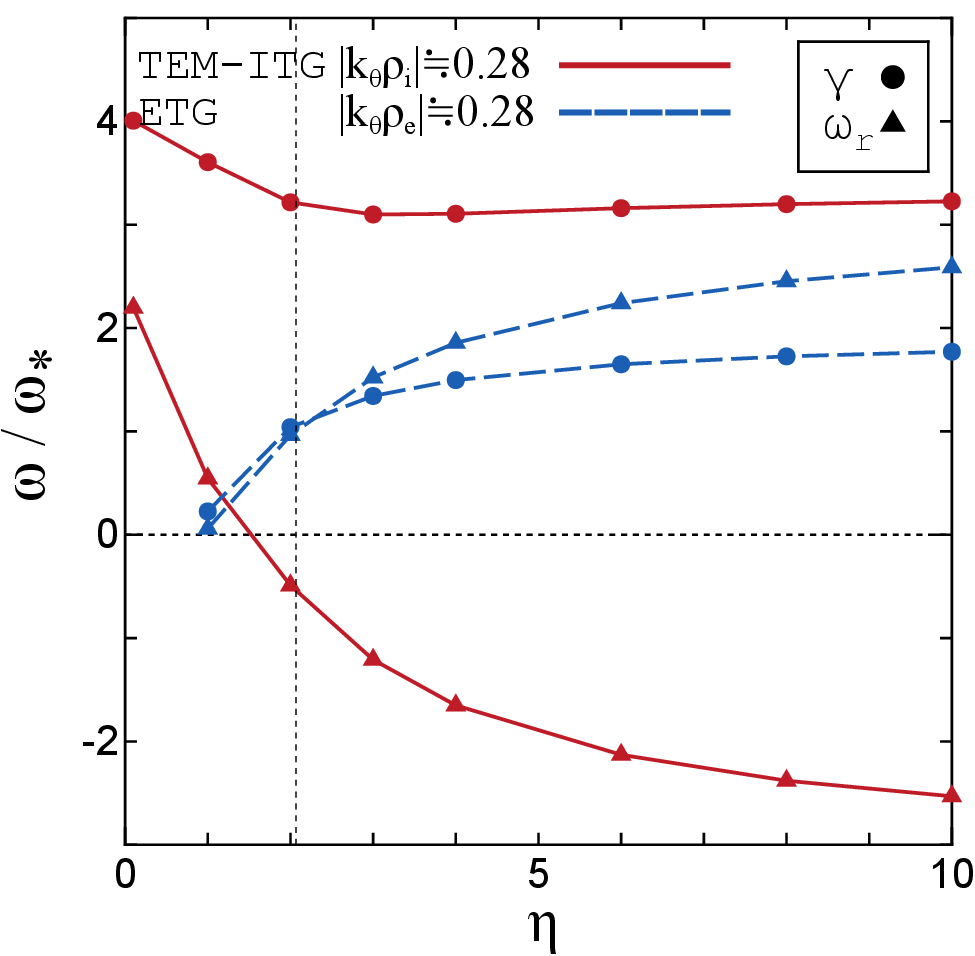}
  \end{minipage}
  \begin{minipage}{.45\linewidth}
    \includegraphics[width=\linewidth]{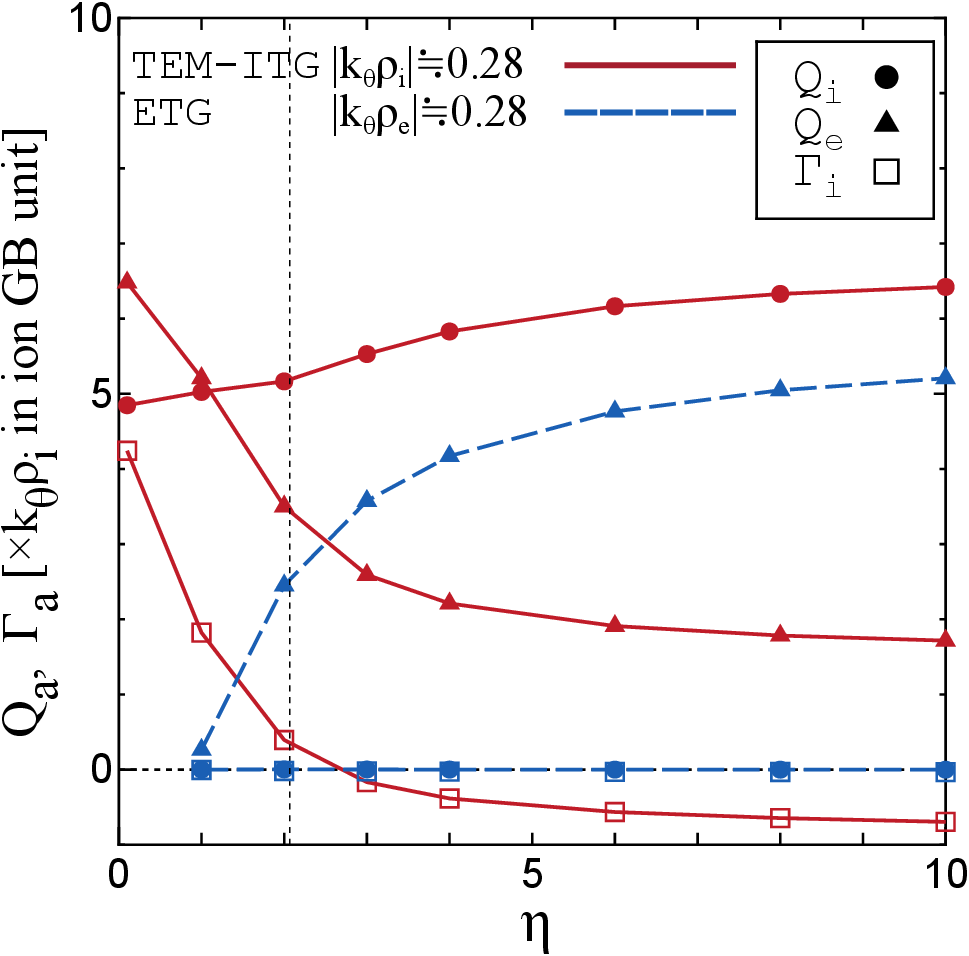}
  \end{minipage}
  \caption{Modulation of instabilities and corresponding fluxes associated with variations in the ratio of temperature to density gradients, $\eta_a = L_{n_a}/L_{T_a}$ ($\eta_i = \eta_e = \eta$),
    where the total gradient $\sum_a(R_0/L_{n_a}+R_0/L_{T_a})$ is kept unchanged from the original case in Fig.\ref{fig:03}, while $\eta$ is varied; the original value, $\eta \approx 2.07$, is indicated by a vertical dotted line.
    (Left) Growth rates (circles) and real frequencies (triangles) of modes with $k_\theta \rho_a \simeq 0.28$
    for ion ($a = i$) and electron ($a = e$) scales, normalized by
    the diamagnetic frequency $\omega_\ast$.
    (Right) Corresponding weighted QL ion (circles) and electron (triangles) energy fluxes $Q_a^k|k_\theta\rho_i|$
    normalized to $p\chi_\textrm{gB}^i/R_0$, along with particle fluxes (squares) $\Gamma_a^k|k_\theta\rho_i|$
    normalized to $n_eD_\textrm{gB}^i/R_0$, where $\Gamma_i=\Gamma_e$ by ambipolarity $\sum_a e_a\Gamma_a\equiv 0$.}
\label{fig:04}
\end{center}
\end{figure}

To examine the QL flux associated with each instability in Fig.~\ref{fig:03},
Fig.~\ref{fig:04} shows the dependence of ion- and electron-scale modes on the ratio of temperature to density gradients,
$\eta_a = L_{n_a}/L_{T_a}$~($\eta_i=\eta_e=\eta$ is assumed below),
where the total gradient at the local position is held fixed while $\eta$ is varied.
The analysis is performed at a representative wavenumber, $|k_\theta \rho_a| = 0.4/\sqrt{2} \simeq 0.28$ ($a=i,e$).
For $\eta \gtrsim 1$,
the ITG mode at $|k_\theta \rho_i| \simeq 0.28$ and the ETG mode at $|k_\theta \rho_e| \simeq 0.28$ are dominant.
The normalized real frequencies shown in Fig.~\ref{fig:04}(left) exhibit approximate mirror symmetry about zero,
consistent with their nearly isomorphic relationship. In contrast, the normalized growth rates differ significantly,
indicating that a trapped-electron contribution persists in the ITG branch even for $\eta \gg 1$. As $\eta$ decreases below unity,
the ETG mode is stabilized, whereas the ion-scale mode remains unstable with a reversal in the sign of the real frequency,
suggesting a transition from ITG to TEM.

The corresponding weighted QL fluxes are shown in Fig.~\ref{fig:04}(right).
For the ETG mode, only $Q_e^{k(\mathrm{ETG})}$ is appreciable,
while $\Gamma_i^{k(\mathrm{ETG})} \sim Q_i^{k(\mathrm{ETG})} \sim 0$, reflecting the approximately adiabatic ion response.
For ITG-dominated modes at $\eta \gg 1$, the particle flux is typically inward, $\Gamma_i^{k(\mathrm{ITG})} < 0$,
with a significant electron energy flux such that $Q_e^{k(\mathrm{ITG})}/Q_i^{k(\mathrm{ITG})} \simeq 1/4$.
The particle flux varies from a large outward value in the TEM-dominated regime ($\eta \ll 1$),
where $\Gamma_i^{k(\mathrm{TEM})} \sim Q_i^{k(\mathrm{TEM})} \sim Q_e^{k(\mathrm{TEM})}$,
to an inward pinch in the ITG-dominated regime. These trends are consistent with nonlinear simulations~\cite{Jenko05,Kinsey07}.
For $\eta\gg 1$,
the weighted QL ion flux from ITG modes and the electron energy flux from ETG modes peak at different wavenumbers,
with $[Q_i^{k(\mathrm{ITG})}]_{|k_\theta\rho_i|=k_0} \gtrsim [Q_e^{k(\mathrm{ETG})}]_{|k_\theta\rho_e|=k_0}$,
where $k_0$($\simeq 0.28$ here) is a numerical value close to the normalized wavenumber at which the growth rates
of ITG and ETG attain their maximum.
This reflects the generally larger normalized growth rate of ITG modes,
which are coupled to trapped electron dynamics, compared with ETG modes,
for which the adiabatic-ion approximation is typically valid.
On the other hand,
a substantial ITG-driven electron energy flux, $Q_e^{k(\mathrm{ITG})}$, emerges,
which is due to the role played by the non-adiabatic electrons.
If these contributions combine additively, the total fluxes satisfy
$[Q_i^{k(\mathrm{ITG})}]_{|k_\theta\rho_i|=k_0}
\simeq[Q_e^{k(\mathrm{ETG})}]_{|k_\theta\rho_e|=k_0} + [Q_e^{k(\mathrm{ITG})}]_{|k_\theta\rho_i|=k_0}$.
Although not shown here,
TEM can persist independently as a secondary instability even in the ITG-dominated regime ($\eta \gtrsim 1$), as in Fig.~\ref{fig:01}.
However, since $Q_i^{k(\mathrm{TEM})} \sim Q_e^{k(\mathrm{TEM})}$,
its contribution does not significantly modify the ion-electron flux balance.
As shown in the lower panel of Fig.~\ref{fig:03}, corresponding to $\eta \approx 2.07$
indicated by the vertical dotted line in Fig.~\ref{fig:04},
the log-linear spectra at ion and electron scales exhibit comparable widths contributing to the area integral for each instability.
This suggests that contributions, arising from well-separated regions in wavenumber space,
add up to yield comparable integrated fluxes, $Q_i \sim Q_e$. As seen in Fig.~\ref{fig:04} (right),
the balance of QL fluxes driven by ITG and ETG,
$[Q_i^{k(\textrm{ITG})}]_{|k_\theta\rho_i|=k_0} \simeq \sum_{a=\textrm{i,e}}[Q_e^{k(\textrm{aTG})}]_{|k_\theta\rho_a|=k_0}$,
is largely insensitive to $\eta$ when the ion and electron temperature gradients are equal,
implying that $Q_i \sim Q_e$ is also a typical trend.
These constitute conclusions closed within the QL flux model; however,
if the above ansatz of area-integral conservation in the nonlinear process holds,
it can be used to interpret the nonlinear simulation results.
Early multiscale simulations, with limited electron-scale resolution,
predominantly captured ion-scale contributions and often yielded $Q_i \gtrsim Q_e$~\cite{Kinsey07,Waltz07,Gorler08},
whereas more recent simulations with sufficient resolution tend to find $Q_i \sim Q_e$~\cite{Holland17,Howard21}.
On the other hand, the predictive capability of the present QL model is expected to break down
when nonlinear effects are strong; this corresponds to cases where nonlinear or externally driven flows reduce ion-scale transport,
leading to $Q_i \lesssim Q_e$~\cite{Holland17,Creely18}.

\section{Conclusions}\label{sec:conclusion}

To summarize, we propose a QL transport flux model [Eq.~(\ref{eq:QL_flux2})]
that incorporates an amplitude prescription consistent with multiscale gyrokinetic ordering
and is self-contained within a linear framework without relying on the nonlinear calibration.
The resulting flux, weighted by $|k_\theta \rho_i|$, is expressed in ion gyro-Bohm units,
making it suitable for log-linear scale representations~\cite{Gorler08,Howard14}.
The amplitude assumption [Eq.~(\ref{eq:model2})] is applied to the electrostatic potential $\phi$
and can be straightforwardly extended to electromagnetic gyrokinetic systems.
In ITG-ETG competitive plasmas,
the ion-scale QL fluxes driven by ITG exhibit wavenumber dependence and magnitude consistent with nonlinear simulations.
The electron energy flux driven by ETG primarily peaks at electron scales, whereas the downshift of the flux toward ion scales,
likely resulting from nonlinear cascade processes,
is not captured by the present model.
Within the framework of the QL model, we obtain the closed result $Q_i \sim Q_e$
due to different instabilities arising at different scales;
this result can also be applied to simulation results if the area-integrated flux is conserved during the nonlinear cascade process.
The result is physically plausible because both species contribute comparably to the energy injection term
in the free-energy balance equation in the absence of toroidal rotation,
$\partial W(r)/\partial t = D + Q$~\cite{Abel13}, for a system with equal temperatures and equal temperature gradients.
Here, $D$ denotes energy dissipation and $Q = \sum_a[\Gamma_a T_a/L_{n_a} + Q_a/L_{T_a}]$ represents the energy injection.
A quantitative assessment of these arguments requires direct comparison with nonlinear simulations,
which remains an important subject for future work.

\begin{acknowledgements}
  The computations in this work were partly performed on Plasma Simulator (NEC SX-Aurora TSUBASA) of NIFS with the support and under the auspices of the NIFS Collaboration Research Program (Grant Nos. NIFS22KIST037, NIFS26KIIP036).
  GW was supported by the National Natural Science Foundation of China (Grant No. 12375039).
\end{acknowledgements}


\end{document}